\documentclass[twocolumn,showpacs,preprintnumbers,amsmath,amssymb]{revtex4}
\usepackage{graphicx}
\usepackage{epsfig}
\usepackage{dcolumn}
\usepackage{bm}

\newcommand{\psip}{\psi(2S)}
\newcommand{\jpsi}{J/\psi}
\newcommand{\rar}{\rightarrow}
\newcommand{\rt}{\rightarrow}

\def\Journal#1&#2&#3(#4){#1{\bf #2}, #3 (#4)}

\def\NPB{Nucl.  Phys.  {\bf B}}
\def\PLB{Phys.  Lett.  {\bf B}}
\def\PRL{Phys.  Rev.  Lett.  }
\def\PRD{Phys.  Rev.  {\bf D}}

\def\etal{{\it et al.}}

\def\bec{\begin{center}}
\def\eec{\end{center}}

\begin{document}
%\title{ $\psi(2S)\rightarrow Vector + Tensor$ }
\title{\boldmath Measurements of $\psi(2S)$ decays into  Vector- Tensor final states}
\author{J.~Z.~Bai$^1$,        Y.~Ban$^{10}$,         J.~G.~Bian$^1$,
X.~Cai$^{1}$,         J.~F.~Chang$^1$,       H.~F.~Chen$^{16}$,    
H.~S.~Chen$^1$,       H.~X.~Chen$^{1}$,      J.~Chen$^{1}$,        
J.~C.~Chen$^1$
% ...
\\(BES Collaboration)\\ 
\vspace{0.2cm}
$^1$ Institute of High Energy Physics, Beijing 100039, People's Republic of
     China\\
$^2$ China Center of Advanced Science and Technology, Beijing 100080,
     People's Republic of China\\
$^3$ Guangxi Normal University, Guilin 541004, People's Republic of China\\
$^4$ Henan Normal University, Xinxiang 453002, People's Republic of China\\
$^5$ Huazhong Normal University, Wuhan 430079, People's Republic of China\\
$^6$ Hunan University, Changsha 410082, People's Republic of China\\                                                  
$^7$ Liaoning University, Shenyang 110036, People's Republic of China\\
$^{8}$ Nanjing Normal University, Nanjing 210097, People's Republic of China\\
$^{9}$ Nankai University, Tianjin 300071, People's Republic of China\\
$^{10}$ Peking University, Beijing 100871, People's Republic of China\\
$^{11}$ Shandong University, Jinan 250100, People's Republic of China\\
$^{12}$ Shanghai Jiaotong University, Shanghai 200030, 
        People's Republic of China\\
$^{13}$ Sichuan University, Chengdu 610064,
        People's Republic of China\\                                    
$^{14}$ Tsinghua University, Beijing 100084, 
        People's Republic of China\\
$^{15}$ University of Hawaii, Honolulu, Hawaii 96822\\
$^{16}$ University of Science and Technology of China, Hefei 230026,
        People's Republic of China\\
$^{17}$ Wuhan University, Wuhan 430072, People's Republic of China\\
$^{18}$ Zhejiang University, Hangzhou 310028, People's Republic of China\\
\vspace{0.4cm}
$^{\ast}$ Visiting professor to University of Michigan, Ann Arbor, MI 48109 USA 
}

\begin{abstract}
Decays of the $\psi(2S)$ into vector plus tensor meson final states
have been studied with 14 million $\psi(2S)$ events collected with the
BESII detector. Branching fractions of $\psi(2S) \rt \omega
f_{2}(1270)$, $\rho a_2(1320)$,
$K^*(892)^0\overline{K}^*_2(1430)^0+c.c.$ and $\phi
f_2^{\prime}(1525)$ are determined.  They improve upon previous BESI
results and confirm the violation of the "12\%" rule for $\psi(2S)$
decays to VT channels with higher precision.
\end{abstract}
\maketitle

%\clearpage
\section{Introduction}
%One of the most mysterious problems confronting the understanding of
%hadronic charmonium decays is the strong suppression of $\psip\rar\rho\pi$
%and $K^*(892) \overline{K}+c.c.$ decays--referred as the $\rho\pi$ puzzle~\cite{rhopi,ichep97}. 
%
%In perturbative QCD, the charmonium states, $J/\psi$ and $\psi(2S)$,
%are considered to be non-relativistic bound states of charm and
%anticharm quarks, and their decays into light hadrons are expected to be
%dominated by the annihilation of the constituent $c$ and $\overline{c}$ quarks
%into three gluons. In this simple picture,
%the partial width for decays into any exclusive hadronic state $h$
% is proportional to the wave function at the origin squared,
%$|\psi(0)|^2$, which is well determined from dilepton decays.
%Since the strong coupling constant $\alpha_s$ does not change much
%between the $J/\psi$ and $\psip$ masses,
%it is reasonable to expect that, for any
%exclusive hadronic state $h$, the $\jpsi$ and $\psip$ decay branching
%fractions will scale as \cite{qcd15}
In perturbative QCD,  the $\jpsi$ and $\psip$ decay branching
 fractions to the same final state are expected to satisfy \cite{qcd15}

\begin{eqnarray*}
Q_h= \frac{B(\psip\rar h)}{B(\jpsi\rar h)}
   \simeq\frac{B(\psip\rar e^+e^-)}{B(\jpsi\rar e^+e^-)}\simeq 12\%,
\end{eqnarray*}
where the leptonic branching fractions are taken from the PDG
tables \cite{PDG}.
This prediction is sometimes referred to as the ``$12\%$ rule''. Although
it seems to work reasonably well for a number of
specific decay modes, it fails severely in the case of the
$\psip$ two-body decays to the vector-pseudoscalar ($VP$) meson final
states, $\rho\pi$ and $K^*\overline{K}$, 
which is the well known ``$\rho\pi$ puzzle"~\cite{rhopi,ichep97}.

Previous BESI results~\cite{BESVT, BES_wphi} on vector-tensor meson
($\omega f_{2}(1270)$, $\rho a_2(1320)$,
$K^*(892)^0\overline{K}^*_2(1430)^0+c.c.$ and $\phi
f_2^{\prime}(1525)$) final states reveal that these VT decay modes are
also suppressed compared to the perturbative QCD prediction.  However,
the measurements, using about $4\times 10^6$ $\psi(2S)$ events,
determined only upper limits or branching fractions with large
errors. Therefore it is hard to tell how strongly these decays are
suppressed with respect to the 12\% rule expectation. Here, we report
the measurement of the branching fractions of $\psi(2S)$ decays into
these four channels with higher precision, based on $14.0\times
10^6(1.00\pm0.04)$ $\psi(2S)$ events~\cite{N_psip} taken with the
upgraded BESII detector.  The results improve on the previous BESI
measurements and confirm the violation of the ``12\%" rule for
$\psi(2S)$ decays to VT channels.

%It is of interest to see if vector-tensor
%($VT$) decay modes are suppressed to the same extent as the hadronic
%$\rho\pi$ and $K^*\bar K$ decays.
%While the previous BESI results~\cite{BESVT} only determined upper limits
%on their branching fractions.
%Based on $14.0\times 10^6(1.00\pm0.04)$ $\psi(2S)$ data~\cite{N_psip} taken in 2001-2002 running 
%year using updated BESII detector, we present the branching fractions of
%$\psi(2S)$ decays into a vector and a tensor mesons with higher precision,
%including  $\omega f_{2}(1270)$, $\rho a_2(1320)$,
%$K^*(892)^0\overline{K}^*_2(1430)^0+c.c.$ 
%and $\phi f_2^{\prime}(1525)$.
% The results update previous
%BESI results and confirm the violation of "12\%" rule for VT channels.

\section{THE BESII DETECTOR}
The Beijing Spectrometer (BESII) is a conventional cylindrical
magnetic detector that is described in detail in Ref.~\cite{BES-II}.
A 12-layer Vertex Chamber (VC) surrounding the beryllium beam pipe
provides input to the event trigger, as well as coordinate
information.  A forty-layer main drift chamber (MDC) located just
outside the VC yields precise measurements of charged particle
trajectories with a solid angle coverage of $85\%$ of $4\pi$; it also
provides ionization energy loss ($dE/dx$) measurements which are used
for particle identification.  Momentum resolution of
$1.7\%\sqrt{1+p^2}$ (p in GeV/$c$) and $dE/dx$ resolution for hadron
tracks of $\sim8\%$ are obtained.  An array of 48 scintillation
counters surrounding the MDC measures the time of flight (TOF) of
charged particles with a resolution of about 200 ps for hadrons.
Outside the TOF counters, a 12 radiation length, lead-gas barrel
shower counter (BSC), operating in limited streamer mode, measures the
energies of electrons and photons over $80\%$ of the total solid angle
with an energy resolution of $\sigma_E/E=0.22/\sqrt{E}$ (E in GeV).  A
solenoidal magnet outside the BSC provides a 0.4 T magnetic field in
the central tracking region of the detector. Three double-layer muon
counters instrument the magnet flux return and serve to identify muons
with momentum greater than 500 MeV/$c$. They cover $68\%$ of the total
solid angle.
%with $z$ ($r \phi$) spatial
%resolution of 5.5 cm (3 cm).

   In this analysis, a GEANT3
based Monte Carlo package (SIMBES) with detailed
consideration of the detector performance (such as dead
electronic channels) is used.
The consistency between data and Monte Carlo has been carefully checked in
many high purity physics channels, and
the agreement is reasonable.
%Phase space generator(HOWL) is used to determine efficiencies.

\section{Event Selection}
The data sample used for this analysis consists of 14 million $\psip$
events, collected with BESII at the center-of-mass energy $\sqrt
s=M_{\psip}$.  The decay channels investigated are $\psip\rar \omega
f_{2}(1270) \rightarrow \pi^+\pi^-\pi^+\pi^-\pi^0$, $\rho
a_2(1320) \rightarrow \pi^+\pi^-\pi^+\pi^-\pi^0$,
$K^*(892)^0\overline{K}^*_2(1430)^0+c.c. \rt \pi^+\pi^-K^+K^-$ and
$\phi f_2^{\prime}(1525) \rt K^+K^-K^+K^-$.
%Hence candidate events must have
%four charged hadrons plus two photons for the
%channels containing a $\pi^0$.
Candidate events are required to
satisfy the following general selection criteria:
\renewcommand{\labelenumi}{\roman{enumi})}
\begin{enumerate}
 \item The number of charged particles must be equal to four with net charge 
     zero.
 \item The number of photon candidates must be equal to or greater than two 
     for the decay channels containing a $\pi^0$.  
 \item For each charged track in an event, the $\chi^2_{PID}(i)$ and 
its corresponding $Prob_{PID}(i)$ values are calculated based on the  $dE/dx$
measurements in the MDC and the TOF measurements in the TOF system, where 
$$\chi^{2}_{PID}(i)=\chi^{2}_{dE/dx}(i)+\chi^{2}_{TOF}(i)$$
$$Prob_{PID}(i)=Prob(\chi^{2}_{PID}(i),ndf_{PID}),$$
where $ndf_{PID}=2$ is the number of degrees of freedom in the $\chi^{2}_{PID}(i)$
determination and $Prob_{PID}(i)$ signifies the probability of this track 
being of particle type $i$ ($i=\pi/K/p$).
For an event to be selected for any signal channel, each track must be
consistent with the expected particle type ($\pi$ or $K$)
by requiring its
%; a track is consistent if 
$Prob_{PID}$ is greater than 0.01 or greater than those for any other assignment.

 \item 
Energy-Momentum conservation is used to provide a four constraint or
five constraint (where the invariant mass of the two photons is also
constrainted to the $\pi^0$ mass for events with a $\pi^0$) kinematic fit
($\chi^{2}_{kine}$) for each event.
To be selected for a candidate final state, the fit probability
must be greater than 0.01.
 \item The combined $\chi^2$, $\chi_{com}^{2}$, is defined as the sum
of the $\chi^2$ values of the kinematic fit ($\chi^{2}_{kine}$) and those from
each of the four particle identification assignments: 
$$\chi_{com}^{2}=\sum_{i}\chi^{2}_{PID}(i)+\chi^{2}_{kine},$$
which corresponds to the combined probability:
$$Prob_{com}=Prob(\chi_{com}^2,ndf_{com}),$$
where $ndf_{com}$ is the corresponding total number of degrees of
freedom in the $\chi_{com}^{2}$ determination.
The final state with the largest $Prob_{com}$ is taken as the  
candidate assignment for each event.
% \item A cut on $R_{Ep}$ is imposed to reject possible contamination 
%  from $\psip\rightarrow\pi^+\pi^-\jpsi$ and $\eta\jpsi$, with  $\jpsi\rar e^+e^-$, where
%  $$R_{Ep}=(\frac{E_{sc}^+}{p_+}-1)^2+(\frac{E_{sc}^-}{p_-}-1)^2,$$
%  and  $p_+$ ($p_-$)is the momentum of positive (negative) charged track measured with the 
%  MDC, and $E_{sc}^+$ ($E_{sc}^-$) is the energy deposited in the BSC by
%  the positive (negative) charged track. 
% \item Hit information from the muon chambers is used to reject possible 
%  muon tracks to reduce contamination from 
%  $\psip\rightarrow\pi^+\pi^-\jpsi$ and $\eta\jpsi$, where $\jpsi\rar\mu^+\mu^-$.   

\item  Backgrounds from $\psi(2S)\rightarrow\pi^{+}\pi^{-}J/\psi$, 
$J/\psi\rightarrow X$ are removed by the $\pi^+\pi^-$ pair recoiling mass requirement:
\begin{eqnarray*}
m_{recoil}^{\pi\pi}
 &= & \sqrt{(E_{cm}-E_{+}-E_{-})^{2}-(\vec{p}_{+}+\vec{p}_{-})^2}  \\
 &\not\in& (3.05,3.15) \hbox{GeV/$c^2$}.
\end{eqnarray*}
where $E_+ (E_-)$ and $\vec{p}_+ (\vec{p}_-)$ are the $\pi^+(\pi^-)$ energy and momentum, respectively.
\end{enumerate}

%\clearpage
\subsection{\boldmath $\psi(2S)\rightarrow \omega f_2(1270)$}
\begin{figure}  \centering
   \includegraphics[height=8.cm,width=0.5\textwidth]{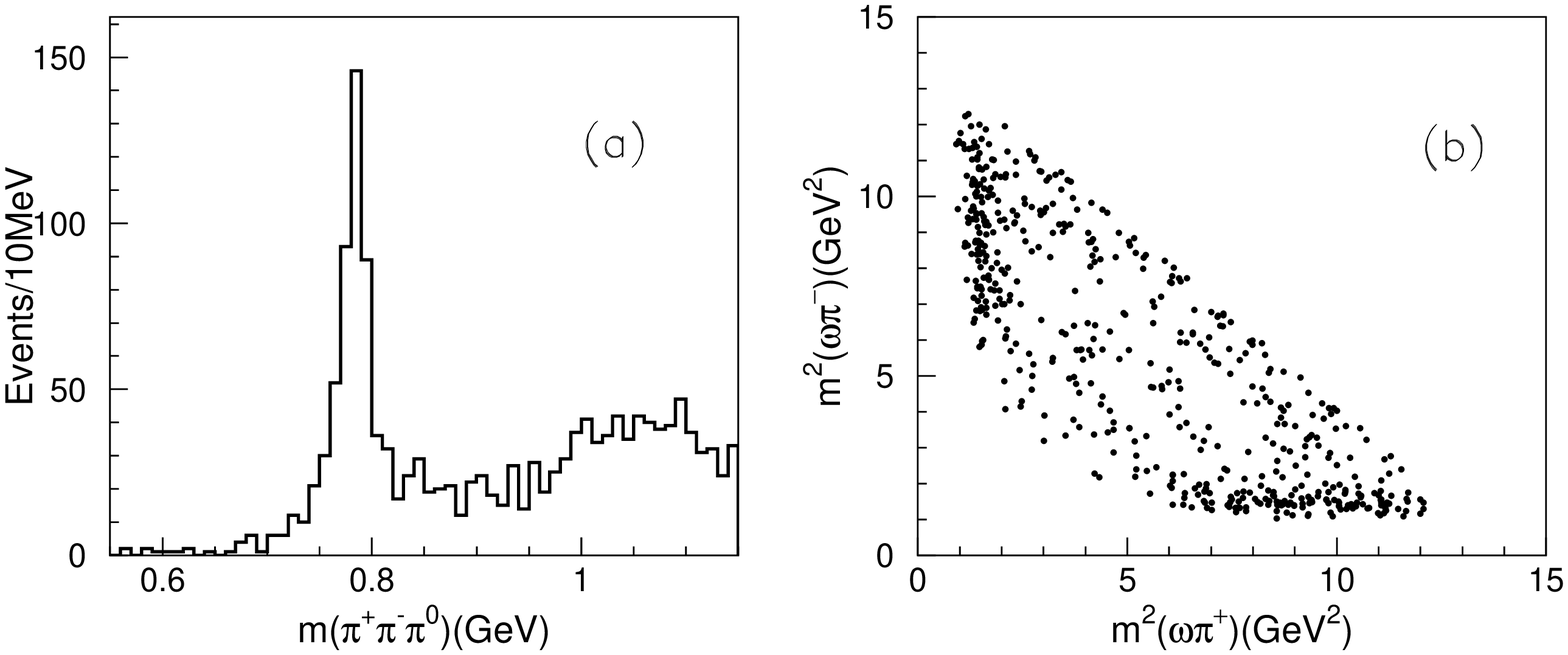}  
  \caption{\label{wpipi}Distributions of $\psi(2S)\rightarrow \omega\pi^+\pi^-$
     candidate events:  (a) the invariant mass of
     $\pi^+\pi^-\pi^0$,  and (b)  the Dalitz plot for
     $\omega\pi^+\pi^-$ events.}
 \end{figure}

The candidate events for this decay mode have the final state
$\pi^+\pi^-\pi^+\pi^-\pi^0$.  To be selected, the combined probability
($Prob_{com}$) for the assignment $\psi(2S)\rightarrow
\pi^+\pi^-\pi^+\pi^-\pi^0$ must be larger than those of
$\psi(2S)\rightarrow \pi^+\pi^-K^+K^-\pi^0$ and $\psi(2S)\rightarrow
\pi^+\pi^-p\overline{p}\pi^0$.  A clear $\omega$ signal is seen in the
$\pi^+ \pi^- \pi^0$ mass distribution, as shown in Fig.~\ref{wpipi}a,
and candidate events are required to satisfy
$|m_{\pi^+\pi^-\pi^0}-0.783|<0.05$ GeV/$c^2$.  An additional
requirement $|m_{\omega\pi^{\pm}}-1.23|>0.2$ GeV/$c^2$ removes almost
all $b_1\pi$ events, which appear as vertical or horizontal bands in
the Dalitz plot shown in Fig.~\ref{wpipi}b.

\begin{figure}[h] \centering
\includegraphics[height=8.cm,width=0.5\textwidth]{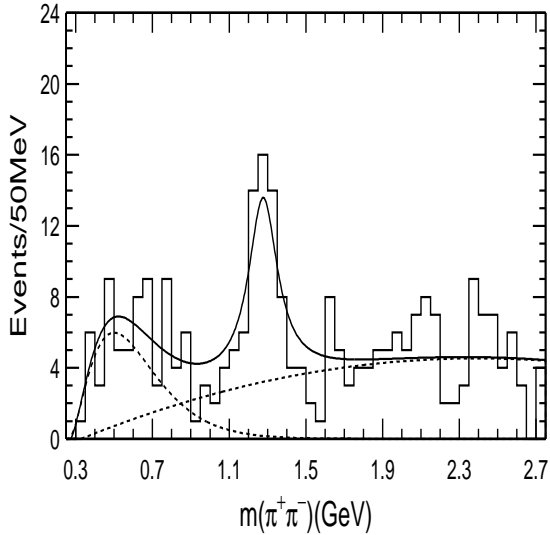}  
\caption{\label{wf2_fit} $M_{\pi^+\pi^-}$ distribution
   of $\omega \pi^+\pi^-$ candidate events. The curves are 
  the results of the fit discussed in the text.}
%  The histogram (dash line) is the backgrounds obtained from both
%  Monte Carlo simulation for $\psi(2S)\rar b_1\pi$ and sidebands of $\omega$.}
\end{figure}

After the above selection, a clear $f_2(1270)$ signal is seen in the
$\pi^+\pi^-$ invariant mass distribution, as shown in
Fig.~\ref{wf2_fit}, along with a smooth background and a broad
enhancement at lower mass, which is attributed to $\sigma$ ($f_{o}(400-1200)$)
production~\cite{sigma}.  Fitting with a Breit-Wigner function for the
$f_2(1270)$ with
mass and width fixed to its PDG
values~\cite{PDG}, plus a second order polynomial for background, and a
$\sigma$, where its spectrum is obtained from $J/\psi$ decays~\cite{sigma},
$62\pm12$ signal events are
obtained.
The statistical significance for the $f_2(1270)$ signal is $6.0\sigma$. 

\subsection{\boldmath $\psi(2S)\rightarrow \rho a_2(1320)$}
\begin{figure}[h] \centering
\includegraphics[height=8.cm,width=0.5\textwidth]{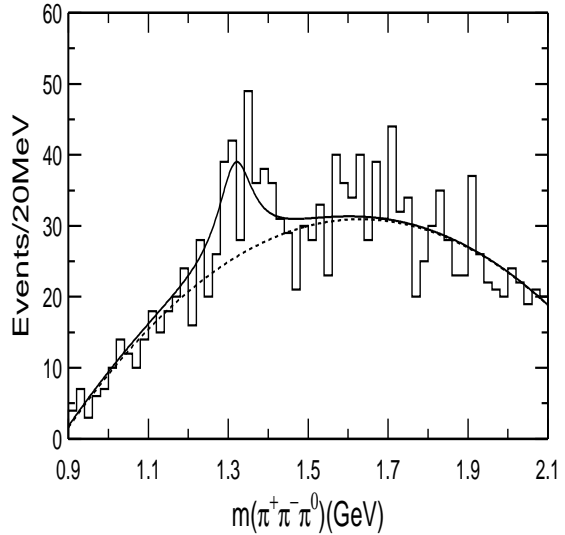}
\caption{\label{rhoa2_fit}
  Distribution of the $\rho\pi$ mass recoiling against another $\rho$.
The  curves are the best fit of the data.}
\end{figure}

The $\pi^+\pi^-\pi^+\pi^-\pi^0$ final state is also used to search for 
$\psi(2S)\rightarrow \rho a_2(1320)\rightarrow \rho\rho\pi$ decay.
Contamination from $\omega\pi^+\pi^-$ is eliminated
by requiring $|m_{\pi^+\pi^-\pi^0}-0.783|>0.03$ GeV/$c^2$.
We select the  $\pi^+\pi^-$ and $\pi^0\pi^{\pm}$ combination
that has the minimum value of 
$\sqrt{(m_{\pi^+\pi^-}-m_{\rho^0})^2+(m_{\pi^0\pi^{\pm}}-m_{\rho^{\pm}})^2}$
and require this minimum value to be less than 200 MeV/$c$.
The combined $\rho^0\pi^{\pm}$ and $\rho^{\pm}\pi^{\mp}$ invariant
mass plot, shown in Fig.~\ref{rhoa2_fit}, has a clear peak near
1320 MeV/$c$.
Assuming the signal is $a_2(1320)$, 
we obtain $112\pm31$ events by fitting the mass distribution with 
a Breit-Wigner function with mass and width fixed at the PDG values~\cite{PDG},
 together with a second order polynomial background function.
The statistical significance  is $3.6\sigma$.

\subsection{\boldmath $\psi(2S)\rightarrow  K^*(892)^0\overline{K}^*_2(1430)^0+c.c.$}
\begin{figure}[h] \centering
\includegraphics[height=8.cm,width=0.5\textwidth]{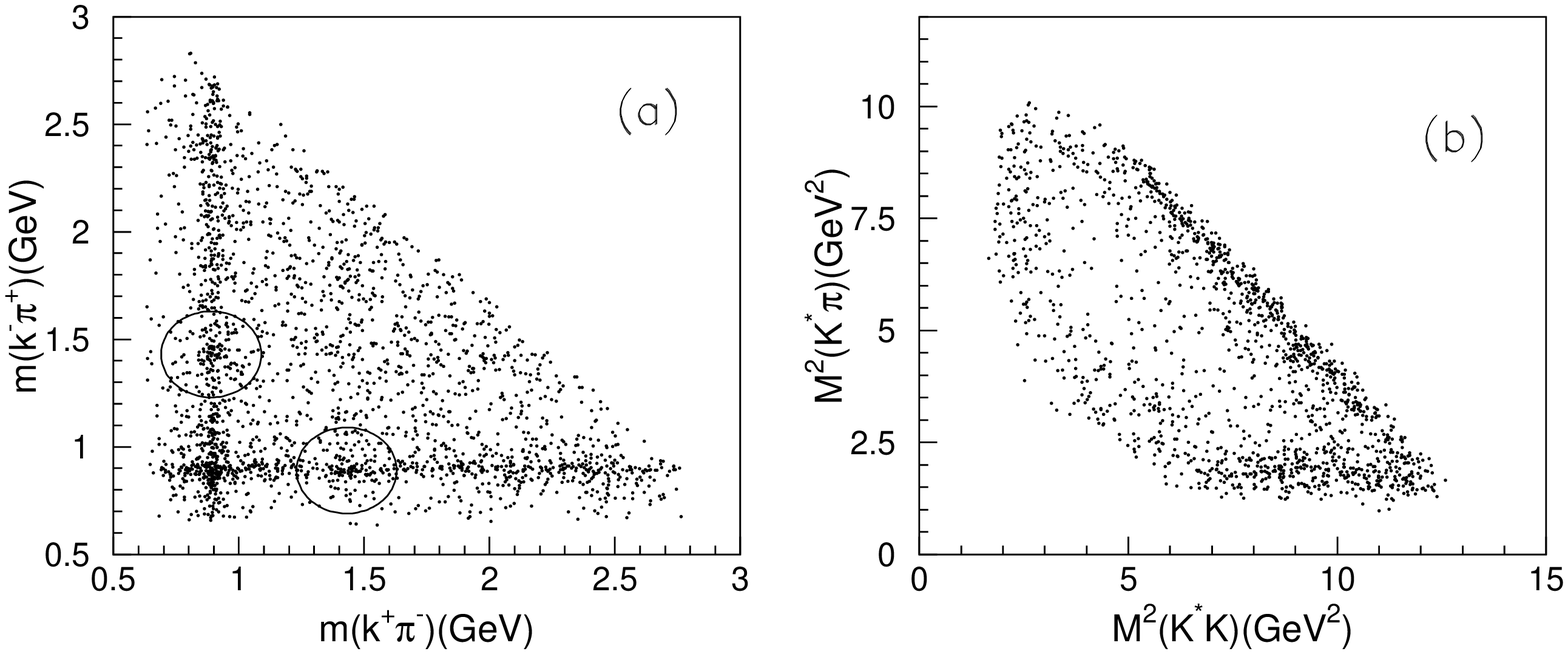}  
\caption{\label{KKpipi}
Distributions of  $\psip\rar K^*(892)K^{\pm}\pi^{\mp}$ candidate events,
(a) scatter plot of $m_{K^+\pi^-}$ versus $m_{K^-\pi^+}$ for selected $\psi(2S)\rightarrow
\pi^+\pi^- K^+K^-$,  and (b) Dalitz plot for $K^*(892)K\pi$ candidate events 
after $K^*(892)$ selection. }
\end{figure}

Candidate events for this decay mode have a final state 
$K^+K^-\pi^+\pi^-$. The combined probability for the assignment of 
$\psi(2S)\rightarrow K^+K^-\pi^+\pi^-$ is required to be larger than those
of $K^+K^-K^+K^-$ and $\pi^+\pi^-\pi^+\pi^-$.
The decay $\psi(2S)\rightarrow\phi \pi^+\pi^-$ is removed by the requirement
 $|m_{K^+K^-}-1.02|>0.02$ GeV/$c^2$.
Candidate  $K^*(892)K^{\pm}\pi^{\mp}$ events are required to satisfy
$|m_{K^{\pm}\pi^{\mp}}-0.896|<0.1$ GeV/$c^2$. The $K\pi$ mass
distribution of these events is shown in Fig.~\ref{KKpipi}a.
We require $m_{\pi^+\pi^-K^{\pm}}>1.6$ GeV/$c^2$ to remove the background
from $K_1(1270)K$, which appears as a horizontal cluster in
Fig.~\ref{KKpipi}b.

\begin{figure}[h] \centering
\includegraphics[height=8.cm,width=0.5\textwidth]{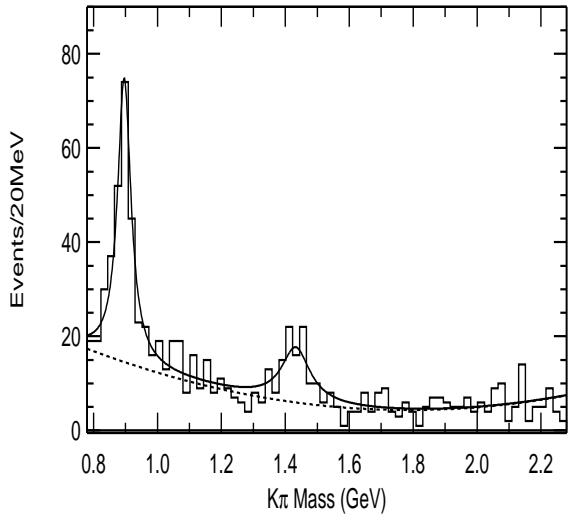}  
\caption{\label{kstar2_fit}Invariant mass of $K\pi$ for 
 $\psi(2S)\rightarrow K^*(892)^0\overline{K}^*_2(1430)+c.c.$ events.
 The curves are the result of the fit described in the text.}
\end{figure}

%After above Xcuts, we can see clear $K^*(892)/\overline{K}^*(892)$ signal and
%$K^*_2(1430)/\overline{K}^*_2(1430)$ signal in the left plot of  Fig.~\ref{KKpipi}.
Fig.~\ref{kstar2_fit} shows a clear peak near $m_{K\pi}=1430$ MeV/$c$.
%Assuming the signal near 1430 MeV/$c$ in Fig.~\ref{KKpipi} almost is 
% $K^*_2(1430)/\overline{K}^*_2(1430)$ events,
By fitting the $K\pi$ invariant mass distribution with two
Breit-Wigner functions
for the $K^*(892)^0$ and $K^*_2(1430)^0$ plus a second order polynomial
background function,  $93\pm16$ events are obtained
with the signal statistical significance of $5.3\sigma$.

%\clearpage
\subsection{\boldmath $\psi(2S)\rightarrow \phi f_2^{\prime}(1525)$}
\begin{figure}[h] \centering
\includegraphics[height=8.cm,width=0.5\textwidth]{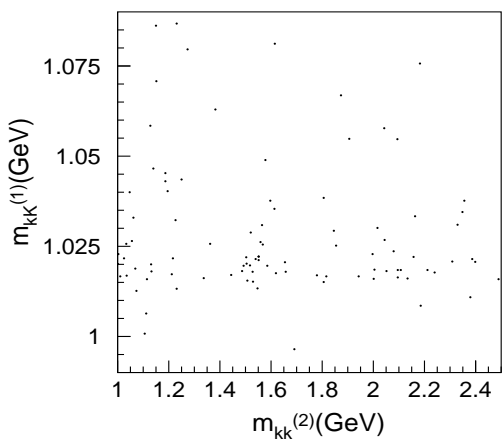}  
\caption{\label{phikk} 
  Evidence for $\psi(2S)\rightarrow \phi f_2^{\prime}(1525)$:
  scatter plot of $m_{K^+K^-}^{(1)}$ versus $m_{K^+K^-}^{(2)}$, where
  the ${K^+K^-}^{(1)}$ pairs is assumed to produced by a $\phi$ and  
  $m_{K^+K^-}^{(2)}$ is the invariant mass recoiling against ${K^+K^-}^{(1)}$. 
  Each event has four entries.}
%   Upper left plot is the invariant mass of $K^+K^-$, upper right is 
%   invariant mass of $K^+K^-$ recoiling against $\phi$( the
%   shadow area is contribution of $\phi$ sideband),
%   the lower  is scatter plot .}
\end{figure}

For this decay, the combined probability for
$\psi(2S)\rightarrow K^+K^- K^+K^-$ is required to be larger than
those of $K^+K^-\pi^+\pi^-$, $K^+K^-p\overline{p}$, and 
$\pi^+\pi^-\pi^+\pi^-$.
 Fig.~\ref{phikk} shows clear evidence for
$\psi(2S)\rightarrow \phi f_2^{\prime}(1525)$.

\begin{figure}[h]
\includegraphics[height=8.cm,width=0.5\textwidth]{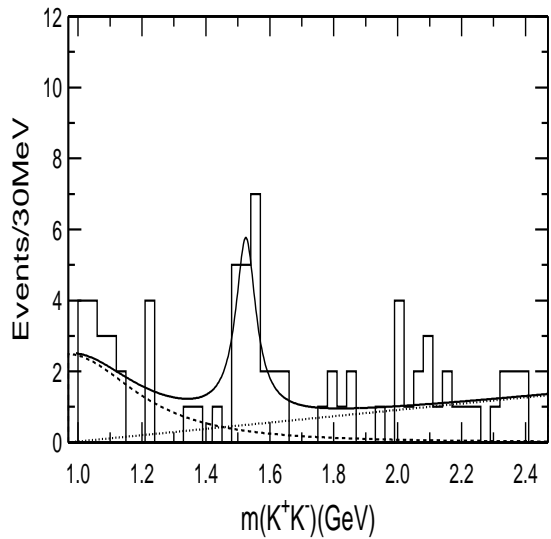}  
\caption{\label{phif2p_fit} Invariant mass distribution of $K^+K^-$
 recoiling against a $\phi$
 for $\phi K^+K^-$ events. The curves are the result of the fit described in the text.}
\end{figure}

Events containing a $\phi$ particle are selected with
the additional requirement $|m_{K^+K^-}-1.02|<0.02$ GeV/$c^2$.
By fitting
the invariant mass $m_{K^+K^-}$ recoiling against a reconstructed $\phi$
particle with a Breit-Wigner function with mass and width of the
 $f_2^{\prime}(1525)$ fixed at its PDG values~\cite{PDG}, plus 
a Flatt\'{e} function for $f_0(980)$ \cite{Flatte} and a first order polynomial for background,
as shown in Fig.~\ref{phif2p_fit}, $19.7\pm5.6$ events are obtained.
The statistical significance of the signal is $4.3\sigma$. 
%$$B=\frac{22.5\pm6.6}{14\times10^6\times0.492\times0.444\times0.137}=(0.54\pm0.16)\times10^{-4} .$$

A possible $\phi f_0(1500)$ state could also decay into
$K^+K^-K^+K^-$, and since the width of $f_0(1500)$ is 109 MeV/$c$, it
could contaminate the $\phi f^{\prime}_2(1525)$ signal.  However, the
branching fraction of $f_0(1500)\rightarrow\pi\pi$ is three times larger
than that of $f_0(1500)\rightarrow K\overline{K}$~\cite{WA102_f0}, and
an analysis of $\psi(2S)\rightarrow \phi
f_0(1500)\rightarrow\phi\pi^+\pi^-$ finds no events from $\phi
f_0(1500)$. Hence, the contamination from
$\psi(2S)\rightarrow\phi K^+K^-$ is neglected.

\section{Systematic errors}
The branching fraction for $\psi(2S)\rightarrow X$ is calculated from
$$B(\psi(2S)\rightarrow X)
  =\frac{n^{obs}_{\psi(2S)\rightarrow X\rightarrow Y}}
   {N_{\psi(2S)}\cdot B(X\rightarrow Y)\cdot \epsilon^{MC}}$$
where X is the intermediate state, Y the final state, and
$\epsilon^{MC}$ the detection efficiency.
Many sources of systematic error are considered. Systematic errors
associated with the efficiency are determined by comparing $J/\psi$
  and $\psi(2S)$ 
data and Monte Carlo simulation for very clean decay channels, such as
  $\psi(2S) \rt \pi^+ \pi^- J/\psi$, which
allows the 
determination of systematic errors associated with
the MDC tracking efficiency, kinematic fitting, particle
identification, and photon selection efficiency \cite{systematics}.

%Since the limited statistical does not allow a determination
%of the angular distribution,
%A phase space generator(HOWL) is used
%to determine efficiencies. The difference between
%HOWL and another generator(UGNT), which generates events depending on 
%particle's spins, parities as well as helicity amplitudes
%invoved in the decay chain, is considered as another source of systematic errors.

Another source of systematic error comes from uncertainties in the
angular distributions used in the simulation.  Events are generated
according to the helicity amplitudes allowed by the spin and parity of
the particles in the decay chain.  However, the limited statistics
does not allow a determination of the helicity amplitudes. This
uncertainty is considered as another source of systematic error.

%\clearpage
\begin{table}[h]  \centering
\caption{\label{TSystematicError}Summary of systematic errors ($\%$).}
%\mbox{} \hskip -1cm
\begin{tabular}{|c|c|c|c|c|c|c|}  \hline\hline \hline
          &$\omega f_2$ & $\rho a_2$ 
	  & $K^{*0}\overline{K}^{*0}_2+c.c.$ 
          &$\phi f_2^{\prime}$ \\ \hline
% mfit  & & & & & \\ \hline
%$|cos\theta|$      & & & & & \\ \hline
Tracking efficiency & \multicolumn{4}{c|}{8.0} \\ \hline
Kinematic fit   &4.0 &4.0 &6.0  &6.0 \\ \hline
PID efficiency      &3.2 &3.2 &4.0 &6.0 \\ \hline
$\gamma$ selection   &5.4 &5.4 & --&-- \\ \hline
MC fluctuation       &2.2 &1.6 &1.3 &1.1 \\ \hline
Helicity &8.1 &1.2  &14.3 &16.0  \\ \hline
Backgrounds shape  &11.0 &13.9 &13.5 & 13.8  \\ \hline
%Sim1040e  &8.0 & 2.7 & 5.9 & 3.7  \\ \hline
Branching fractions  &2.9 &3.4 &2.4 &3.8 \\ \hline
%Contamination & -- &--& -- &2.6 \\ \hline
%Continue contribution &14.4 &24.5 & 7.9 & 29.3 \\ \hline
 $N_{\psi(2S)}$   & \multicolumn{4}{c|}{4.0} \\ \hline
Sum                   &18.3 &18.5 &23.2 &25.0 \\ \hline \hline \hline
\end{tabular}
\end{table}

Contributions from the continuum $e^+ e^- \rightarrow
\gamma^*\rightarrow$ hadrons \cite{wangp} are estimated using a data
sample of $\sim 6.0$ pb$^{-1}$ taken at $\sqrt s=3.65$ GeV/$c^2$,
about one-third of the integrated luminosity at the $\psi(2S)$.  No signal
is found for any channel under study, hence this background is
neglected.  The uncertainties of the branching fractions of intermediate
states, the background shapes, and the total number of $\psi(2S)$
events are also
sources of systematic errors.  Table~\ref{TSystematicError} summarizes
the systematic errors for all channels; the total branching
fraction errors for $\omega f_2$,$\rho a_2$,
$K^{*0}\overline{K}^{*0}_2+c.c$. and $\phi f_2^{\prime}$ are 18.3\%,
18.5\%, 23.2\% and 25.0\%, respectively.

%%%%%%%%%%%%%%%%%%%%%%%%%%%%%%%%%%%%%%%%%%%%%%%%%%%%%%%%
\section{Results}

Table~\ref{BESII_VT} summarizes the results of the four branching
fraction measurements.
For comparison, the table includes the corresponding decay branching fractions of  $J/\psi$
decays~\cite{Jpsi_Bf}, as well as the ratios of the $\psi(2S)$ to $J/\psi$
branching fractions. 
These results have smaller statistical errors than the previous BESI
measurements, mainly due to the larger $\psi(2S)$ event sample.
 The statistical significances for all four channels are larger than
$3 \sigma$; those for $\omega f_2(1270)$ and
$K^*(892)^0\overline{K}^*(1430)^0+c.c.$ are larger than $5 \sigma$.

 In perturbative QCD, VP decays are forbidden by
hadron helicity conservation (HHC)~\cite{hhc1}, whereas VT decay are
HHC allowed~\cite{hhc2}.  Although the suppression of the VT decays is
not as severe as that of the $\rho\pi$ and $K^*\overline{K}$ decay
channels, all four VT decay modes are suppressed by a factor of 3 to 5
compared with the pQCD expectation.

%{\tiny
\begin{table*} \centering
\caption{\label{BESII_VT}
  Branching fractions measured for 
  $\psi(2S)\rightarrow$ Vector + Tensor. Results for corresponding
  $J/\psi$ branching fractions \cite{Jpsi_Bf} are also given as well as the ratio
  $Q_X=\frac{B(\psi(2S)\rar X)}{B(J/\psi\rar X)}$. } \vskip 2pt
\begin{ruledtabular}
\begin{tabular}{|c|c|c|c|c|c|}  \hline \hline
X &$N^{obs}$ & $\epsilon(\%)$ & $B(\psi(2S)\rar X)(\times 10^{-4})$ &
$B(J/\psi\rar X)(\times 10^{-3})$ & $Q_X(\%)$ \\ \hline
$\omega f_{2}$  & $62\pm12$  & $4.25\pm0.10$  & $2.05\pm0.41\pm0.38$
 & $4.3\pm0.6$ & $4.8\pm 1.5$ \\
$\rho a_2$              & $112\pm31$ & $6.42\pm0.06$
& $2.55\pm0.73\pm0.47$ & $10.9\pm2.2$ & $2.3\pm1.1$ \\
$K^*\overline{K}^*_2$   & $93\pm16$ & $16.2\pm0.2$
& $1.86\pm0.32\pm0.43$ & $6.7\pm2.6$  &  $2.8\pm1.3$ \\
$\phi f_2^{\prime}$     & $19.7\pm5.6$  & $14.8\pm0.2$ &
$0.44\pm0.12\pm0.11$
 & $1.23\pm 0.21$ & $3.6\pm1.5$ \\ \hline \hline
\end{tabular} \\
\end{ruledtabular}
\end{table*}

\acknowledgments
   The BES collaboration thanks the staff of BEPC for their 
hard efforts. This work is supported in part by the National 
Natural Science Foundation of China under contracts 
Nos. 19991480, 10225524, 10225525, the Chinese Academy
of Sciences under contract No. KJ 95T-03, the 100 Talents 
Program of CAS under Contract Nos. U-11, U-24, U-25, and 
the Knowledge Innovation Project of CAS under Contract 
Nos. U-602, U-34(IHEP); by the National Natural Science
Foundation of China under Contract No. 10175060 (USTC); 
and by the Department of Energy under Contract 
No. DE-FG03-94ER40833 (U Hawaii).


\begin{thebibliography}{99}
\bibitem{Jpsi_Bf} For the first three decays, we use the PDG recommended
  values of the $J/\psi$ branching fractions. For $J/\psi$ decay
  to $\phi f^{\prime}_2$, there is poor consistency between the existing 
  measurement, Mark II and DM2. Instead of the PDG average with 50\% error,
  we use data of DM2, which was a high statistics measurement,
  giving smaller errors and including the interference with $f_0(1710)$.
\end{thebibliography}
\end{document}